\documentclass[11pt,twoside]{article} 
\usepackage{asp2004}
\usepackage{epsf}
\usepackage{psfig}
\usepackage{lscape}

\begin{document} 
\title{Identification of a DO White Dwarf and a PG1159 Star in the ESO SN~Ia Progenitor Survey (SPY)}

\author{K. Werner,$^1$ T. Rauch,$^{1,2}$ R. Napiwotzki,$^3$
  N. Christlieb,$^4$ D. Reimers,$^4$ and C.A. Karl$^2$}
\affil{$^1$Institut f\"ur Astronomie und Astrophysik, Universit\"at T\"ubingen,
 Sand~1, 72076 T\"ubingen, Germany\\
 $^2$Dr.~Remeis-Sternwarte, Sternwartstr.~7, 96049 Bamberg, Germany\\
 $^3$Department of Physics and Astronomy, University of Leicester, University Road, Leicester, LE1 7RH, UK\\
 $^4$Hamburger Sternwarte, Universit\"at Hamburg, Gojenbergsweg 112, 21029 Hamburg, Germany
}

\begin{abstract}
We present high-resolution VLT spectra of a new DO white dwarf and a new PG1159
star, which we identified in the ESO SPY survey. The PG1159 star is a
low-gravity, extremely hot ($T_\mathrm{eff}=160\,000$\,K, $\log g=6$) star,
having a C/He dominated atmosphere with considerable amounts of O and Ne
(He=38\%, C=54\%, O=6\%, Ne=2\% by mass). It is located within the planetary
nebula nuclei instability strip and pulsations have been discovered. The DO is a
unique object. From \ion{He}{i}/\ion{He}{ii} line strengths  we found
\mbox{$T_\mathrm{eff}$}$\approx$60\,000\,K, however, the \ion{He}{ii} lines are
extraordinarily strong and cannot be fitted by any model.
\end{abstract}

\section{Introduction} 

The ESO Supernovae Ia Progenitor Survey (SPY) is aimed at finding binary WDs  to
test the double-degenerate scenario for SN~Ia progenitors (Napiwotzki et\,al.\
2001). Here we report on the identification and spectrum analysis of a PG1159
star and a DO WD. HE1314+0018 (B=15.6\,mag) and HE1429$-$1209 (B=15.8\,mag)  were
identified in the Hamburg ESO survey (HES; Wisotzki et\,al.\ 2000, Christlieb
et\,al.\ 2001) as WD candidates and were therefore included in the SPY
project. The spectra presented here were taken between April 2000 and July
2002. Their resolution is about $R=18\,500$. Data reduction was performed with
the ESO MIDAS software package.  Line blanketed non-LTE model atmospheres were
computed using our {\sc PRO2} code. The models assume plane-parallel geometry
and hydrostatic and radiative equilibrium.

\section{The DO Star}

The spectrum of HE1314+0018 exhibits lines from neutral and ionized
helium, i.e., we have detected a ``cool'' DO (``hot'' DOs show only
\ion{He}{ii} lines). This usually allows an accurate determination of
\mbox{$T_\mathrm{eff}$} and \mbox{$\log g$}.  We also detect the
\ion{C}{iv}~5801/5812\,\AA\ doublet and a very weak feature of this
ion close to \ion{He}{ii}~4686\,\AA.  Surprisingly, we are unable to
obtain an acceptable fit to the observed spectrum. Fig.\,1 shows the
best fit model to the \ion{He}{i} lines, which is obtained with
\mbox{$T_\mathrm{eff}$}=60\,000\,K and \mbox{$\log g$}=7.5. It is
obvious that the synthetic \ion{He}{ii} lines are much too weak
compared to the observation. We found no model that can fit the strong
\ion{He}{ii} lines.  If we increase \mbox{$T_\mathrm{eff}$}, these
lines become slightly deeper, but never reach the observed equivalent
widths. \mbox{$T_\mathrm{eff}$}$>$70\,000\,K can be excluded because
the \ion{He}{i} lines disappear. \mbox{$T_\mathrm{eff}$}$<$55\,000\,K
is excluded because the \ion{He}{i} lines become too strong and the
\ion{He}{ii} lines are getting even weaker. A compromise is reached at
about 65\,000\,K in a sense that the relative \ion{He}{i}/\ion{He}{ii}
line strengths in the model are similar to the observed ones. But,
clearly, lines from both ionization stages are much too weak.

\begin{figure}
\epsfxsize=\textwidth  \epsffile{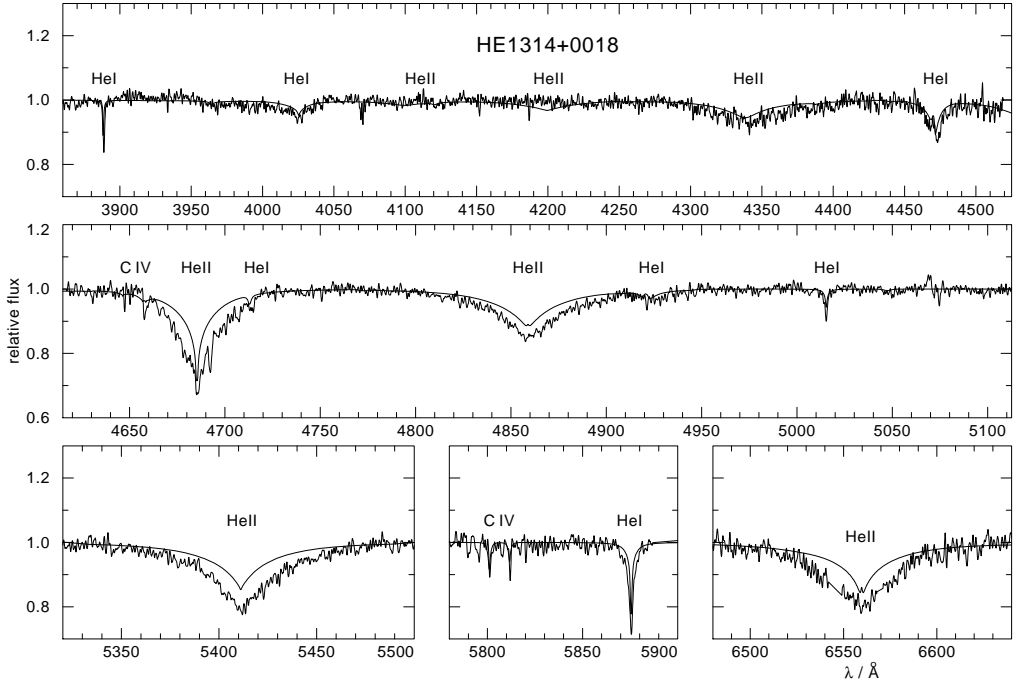}
\caption{Spectrum of the new DO white dwarf. Overplotted is a model with
\mbox{$T_\mathrm{eff}$}=60\,000\,K, \mbox{$\log g$}=7.5, and C/He=0.001. The
\ion{He}{i} lines fit well, but all the \ion{He}{ii} lines in the model are much
too weak. No acceptable fit to the complete spectrum is found.}
\end{figure}

This situation was never encountered before in analyses of cool
DOs. However, this problem is reminiscent of what we faced with those
``hot'' DOs that show signatures of a super-hot wind. We have found
that a large fraction (50\%) of the hot DOs shows such signatures in
the optical spectrum (Werner et\,al.\ 1995). These hot stars show high
ionization absorption lines of the CNO elements (e.g.\ \ion{C}{vi},
\ion{N}{vii}, \ion{O}{viii}, and even \ion{Ne}{x}). The high
excitation potentials involved require temperatures approaching almost
10$^6$\,K and the triangular shaped line profiles suggest their
formation in a rapidly accelerating wind from the WD, reaching a
terminal velocity of the order of 10\,000\,km\,s$^{-1}$. At the same
time, all \ion{He}{ii} lines have symmetric profiles and, like in the
present case, they are much too strong to be fitted by any DO
model. The reason for this simultaneous occurrence of ``hot-wind''
signatures and too-strong \ion{He}{ii} lines in hot DOs is unknown.

We have also considered the possibility that HE1314+0018 shows a binary
composite spectrum. This can be ruled out because, as already mentioned, the
strong \ion{He}{ii} lines cannot be matched by any DO model, even if we
disregard the presence of \ion{He}{i} lines.

From the \ion{C}{iv}~5801/5812\,\AA\ doublet we find a carbon abundance of
C/He$=0.003$ by number, under the assumption of \mbox{$T_\mathrm{eff}$}
=60\,000\,K and \mbox{$\log g$}=8, hence, HE1314+0018 is the coolest DO with a
safe detection of a photospheric trace metal.

\section{The PG1159 Star}

A comparison of the HE1429$-$1209 spectrum with other PG1159 stars immediately
reveals that we have found another very hot, low-gravity PG1159 star which is
rather similar to the well known central star of the planetary nebula NGC\,246
and the central star RX\,J2117.1$+$3412. According to the classification scheme
introduced by Werner (1992), the spectral subtype is ``lgE'' (meaning
low-gravity star with emission lines). The spectrum is characterized by lines
from highly ionized species (\ion{He}{ii}, \ion{C}{iv}, \ion{O}{vi}, Fig.\,2),
often present as absorption lines with central emission reversals or even as
pure emission lines.

We stress in particular the presence of a \ion{Ne}{vii} line at
3644\,\AA\ and a \ion{Ne}{vii} multiplet in the 3850--3910\,\AA\
range. We have recently performed a systematic investigation about
neon in a large number of PG1159 stars (Werner et\,al.\ 2004a) and
found that such prominent \ion{Ne}{vii} lines are preferentially
exhibited by ``lgE'' subtypes. It has been shown that neon is strongly
overabundant (about 20 times solar).  For HE1429$-$1209 we find:
\vspace*{-0.1em}
\begin{eqnarray*}
\mbox{$T_\mathrm{eff}$}   &=& 160\,000\,{\rm K} \pm 15\,000\,{\rm K} \qquad \mbox{$\log g$}=6.0 \pm 0.3 {\rm
\ \ [cm/s}^2{\rm ]}\\ {\rm He}&=&38\% \quad   {\rm C} =54\% \quad    {\rm O}
=6\% \quad   {\rm Ne}=2\%
\end{eqnarray*}
\vspace*{-0.1em}
The estimated error for abundances is 0.3\,dex.  Stellar mass, luminosity, and
distance can be derived by comparing the star's position in the \mbox{$\log
g$}--\mbox{$T_\mathrm{eff}$} diagram with theoretical evolutionary tracks:
\vspace*{-0.1em}
\begin{eqnarray*}
M/{\rm M}_\odot&=&0.68^{+0.15}_{-0.08} \qquad \log L/{\rm
  L}_\odot=4.04^{+0.09}_{-0.05} \qquad  d/{\rm kpc}=5.2^{+1.6}_{-2.2}
\end{eqnarray*}
\vspace*{-0.1em}
HE1429$-$1209 is still very luminous, i.e., on the horizontal part of
the post-AGB evolutionary track in the HRD, located within the domain
of the hottest central stars of planetary nebulae. It is among the
hottest PG1159 stars.  It is very similar to six members (having
subtype lgE), which also are hot high-luminosity objects. It is thus
not surprising that all except one of these previously known
lgE-PG1159 stars have associated planetary nebulae, which are rather
old and large, but not yet dispersed as is the case for most of the
more evolved PG1159 stars (having entered the WD cooling sequence). So
we expect that our new PG1159 star could also be surrounded by an
extended PN.

In the \mbox{$\log g$}--\mbox{$T_\mathrm{eff}$} diagram HE1429$-$1209 is located
within the GW\,Vir instability strip, close to the PNN variables K\,1-16,
RX\,J2117.1$+$3412, NGC\,246, and Longmore~4. And in fact, we could discover
pulsations in the new PG1159 star (Nagel \& Werner 2004). More details on the
work described here were published in A\&A (Werner et\,al.\ 2004b).

\begin{figure}[t]
\epsfxsize=\textwidth \epsffile{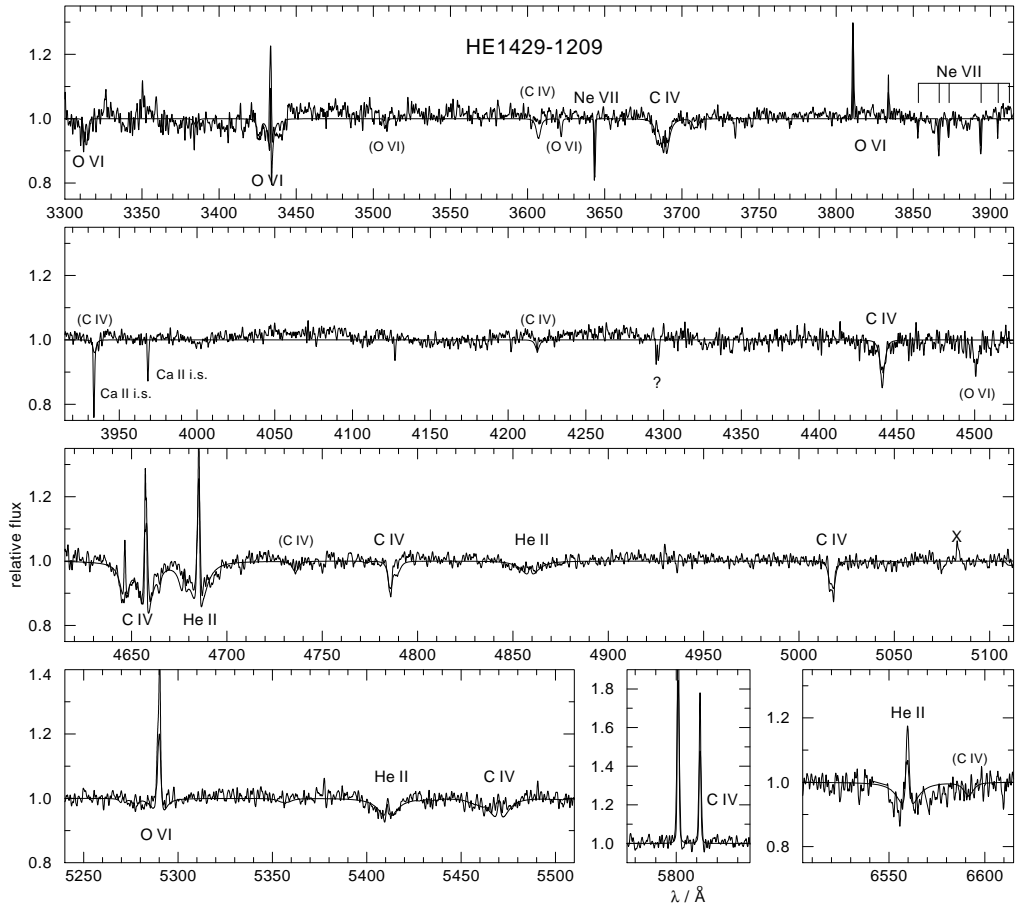}
\caption{Spectrum of the new PG1159 star with the best fit model overplotted. 
Note the discovery of a \ion{Ne}{vii} multiplet in the 3850--3910\,\AA\ range.}
\end{figure}

\acknowledgements{T.R.\ is supported by DLR grant 50\,OR\,0201, C.K.\
by DFG grant NA365/2-2, and R.N.\ by a PPARC advanced fellowship.}

\end{document}